\documentclass[aps,preprint,twocolumn,10pt]{revtex4}
 \pdfoutput=1
\usepackage{hyperref}
\usepackage{amsfonts}
\usepackage{amsmath}

\usepackage{amssymb}
\usepackage{graphicx}%
\usepackage{bigints}
\usepackage{graphicx}
\usepackage{subfigure}
\usepackage{epsf}
\usepackage{bm}
\usepackage{amsmath}
\usepackage{amsfonts}
\usepackage{amssymb}
\usepackage{graphicx}
\usepackage{tabularx}
\usepackage{color}%
\usepackage{multirow}
\providecommand{\U}[1]{\protect\rule{.1in}{.1in}}

\newcommand{\be}{\begin{equation}}
\newcommand{\ee}{\end{equation}}

\newcommand{\mincir}{\raise
-3.truept\hbox{\rlap{\hbox{$\sim$}}\raise4.truept\hbox{$<$}\ }}
\newcommand{\magcir}{\raise
-3.truept\hbox{\rlap{\hbox{$\sim$}}\raise4.truept\hbox{$>$}\ }}

\newcolumntype{Y}{>{\centering\arraybackslash}X}
\providecommand{\U}[1]{\protect\rule{.1in}{.1in}}

\usepackage{tikz,xcolor,hyperref}

\definecolor{lime}{HTML}{A6CE39}
\DeclareRobustCommand{\orcidicon}{%
	\begin{tikzpicture}
	\draw[lime, fill=lime] (0,0) 
	circle [radius=0.16] 
	node[white] {{\fontfamily{qag}\selectfont \tiny ID}};
	\draw[white, fill=white] (-0.0625,0.095) 
	circle [radius=0.007];
	\end{tikzpicture}
	\hspace{-2mm}
}

\foreach \x in {A, ..., Z}{%
	\expandafter\xdef\csname orcid\x\endcsname{\noexpand\href{https://orcid.org/\csname orcidauthor\x\endcsname}{\noexpand\orcidicon}}
}

\begin{document}

\title{\Large{Dark Universe Phenomenology from Yukawa Potential?}}

\author{Kimet Jusufi\orcidA{}$^2$}
\email{kimet.jusufi@unite.edu.mk}
\author{Genly Leon\orcidB{}$^{1,3}$}
\email{genly.leon@ucn.cl}
\author{Alfredo D. Millano\orcidC{}$^1$}
\email{alfredo.millano@alumnos.ucn.cl}
\affiliation{$^1$Departamento de Matem\'{a}ticas, Universidad Cat\'{o}lica del Norte, Avda.
Angamos 0610, Casilla 1280 Antofagasta, Chile}
\affiliation{$^2$Physics Department, State University of Tetovo, 
Ilinden Street nn, 1200, Tetovo, North Macedonia}
\affiliation{$^3$Institute of Systems Science, Durban University of Technology, PO Box 1334, Durban 4000, South Africa}

\begin{abstract}
 We argue that the effect of cold dark matter in the cosmological setup can be explained by the coupling between the baryonic matter particles in terms of the long-range force having a graviton mass $m_g$ via the Yukawa gravitational potential. Such a quantum-corrected Yukawa-like gravitational potential is characterized by the coupling parameter $\alpha$, the wavelength parameter $\lambda$, which is related to the graviton mass via $m_g=\hbar/(\lambda c)$, that determines the range of the force and, finally, a Planck length quantity $l_0$ that makes the potential regular at the centre. The modified Friedmann equations are obtained using Verlinde's entropic force interpretation of gravity based on the holographic scenario and the equipartition law of energy. The parameter $\alpha$ modifies the Newton's constant as $G_{\rm eff} = G \,(1+\alpha)$. Interestingly, we find an equation that relates the dark matter density, dark energy density, and baryonic matter density. It is given by $\Omega_D= \sqrt{2\, \Omega_B  \Omega_{\Lambda}}{(1+z)^3}$. We argue that dark matter is an apparent effect as no dark matter particle exists in this picture. Furthermore, the dark energy is also related to graviton mass and $\alpha$; in particular, we point out that the cosmological constant can be viewed as a self-interaction effect between gravitons. We further show a precise correspondence with Verlinde's emergent gravity theory, and due to the long-range force, the theory can be viewed as a non-local gravity theory. To this end, we performed the phase space analyses and estimated $\lambda \simeq 10^3 $ [Mpc] and $\alpha \in 0.04$, respectively. Finally, from these values, for the graviton mass, we get $m_g \simeq 10^{-68}$ kg, and cosmological constant $\Lambda \simeq 10^{-52} \, \rm m^{-2}$. Further, we argue how this theory reproduces the MOND phenomenology on galactic scales via the acceleration of Milgrom $a_0 \simeq 10^{-10}\,\rm m/s^2$.
\end{abstract}
\maketitle
\section{Introduction}

It is well known that the $\Lambda$CDM cosmological model describes well the evolution and large-scale structure of the Universe. According to this model, the Universe comprises cold dark matter, dark energy, and ordinary matter (baryonic matter) \cite{Springel:2006vs}. Cold dark matter (DM) is a matter that does not interact electromagnetically and can only be detected through its gravitational effects. Among many problems, the cold dark matter is assumed to solve the flat rotation curves in galaxies \cite{Navarro:1996gj, Navarro:2008kc, Moore:1999gc, Gilmore:2007fy, Salucci:2007tm, KuziodeNaray:2007qi}, but as of today, no dark matter particle has been detected.
On the other hand, the cosmological constant ($\Lambda$) is considered a form of energy that permeates the entire Universe, acts as a repulsive form of energy, and is linked to the observed acceleration of the Universe's expansion. The cosmological constant is very often linked to the vacuum energy; however, there is an open problem in cosmology which has to do with the fact that the measured value is approximately 120 orders of magnitude smaller than the value predicted by any quantum gravity theory \cite{Weinberg:1988cp}. The true nature of dark matter and dark energy has yet to be resolved and raises the question about the validity of the entire $\Lambda$CDM framework.

On the other hand, a class of modified gravity theories are proposed to explain the observed phenomena in the Universe, including the origin of dark matter or dark energy. To explain the flat rotation curves, the Modified Newtonian dynamics (MOND) was proposed by Milgrom \cite{Milgrom:1983ca}. This theory modifies Newton's law and explains the flat rotation curves of spiral galaxies \cite{Ferreira:2009eg, Milgrom:2003ui, Tiret:2007kq, Kroupa:2010hf, Cardone:2010ru, Richtler:2011zk, Bekenstein:2004ne}. Other interesting ideas like the superfluid dark matter theory to explain dark matter and reproduce MOND has been suggested \cite{Berezhiani:2015bqa}, and the Bose-Einstein condensate \cite{Boehmer:2007um} has been proposed. In the present paper, we shall follow an approach motivated by cosmology and quantum field theories; we aim to study the dark sector of the Universe by introducing the Yukawa potential \cite{Garny:2015sjg, Arvanitaki:2016xds, Desmond:2018euk, Desmond:2018sdy, Tsai:2021irw}. Furthermore,  we adopt the viewpoint of Verlinde and
consider gravity as an entropic force caused by the changes in the
system's information \cite{Verlinde:2010hp}. Verlinde further argued that dark matter is an apparent effect, i.e., a consequence of the baryonic matter \cite{Verlinde:2016toy}. Furthermore, the entropic force was recently used in deriving the corrected Friedmann equations due to the minimal length \cite{Jusufi:2022mir, Jusufi:2023ayv}. In this paper, we want Friedmann's modified equations due to a quantum-modified Yukawa gravitational potential. Such a Yukawa gravitational potential modifies Newton's potential in short and large distances. In literature, this potential is studied widely for various cosmological scenarios. For instance, Zurab Berezhiani et al. \cite{Berezhiani:2009kv} have presented that the Newtonian gravitational potential is possible to modify through Yukawa modification at a linear level which opens up the possibility of having interaction between dark matter and ordinary matter in a physically non-trivial way through modified gravitational interaction \cite{Berezhiani:2009kv}. Such a potential is interesting since it modifies Newton's potential in short and large distances.\\

A complementary approach uses a dynamical systems analysis to determine asymptotic states/solutions  \cite{wainwright1997dynamical}. This study consists of several steps: determining singular points, the linearization in a neighbourhood of them, the search for the eigenvalues of the associated Jacobian matrix, checking the stability conditions in a neighbourhood of the singular points, the finding of the stability and instability sets and the determination of the basin of attraction, etcetera. Dynamical systems tools and observational tests have been explored and applied in several cosmological contexts \cite{Ovgun:2017iwg, Hernandez-Almada:2020uyr, Paliathanasis:2020abu, Hernandez-Almada:2021rjs, Hernandez-Almada:2021aiw, Leon:2021wyx, Garcia-Aspeitia:2022uxz, Gonzalez:2023who}. These methods have been proven to be a robust scheme for investigating the physical behaviour of cosmological models and can be applied in new contexts, as in this proposal.

This paper is outlined as follows. In Section \ref{Sect:II}, we derive the
corrected entropy and the modified Friedmann equations with Yukawa potential. Moreover, we study the phase space analyses in Section \ref{Sect:III}. Finally, we comment on our results in Section \ref{Sect:IV}.

\section{Entropic Corrections to Friedmann Equations via Yukawa Potential}
\label{Sect:II}
In the present paper, we will use Verlinde's entropic force scenario, according to which when a test particle or
excitation moves apart from the holographic screen, the magnitude
of the entropic force on this body has the form \cite{Verlinde:2010hp}
$F\triangle x=T \triangle S$. 
Note here that $\triangle x$ gives the displacement of the particle from the
holographic screen, while $T$ and $\triangle S$ are the
temperatures and the entropy change on the screen, respectively. Another important point in Verlinde's derivation of Newton's law of gravitation is the entropy-area relationship $S=A/4$ of
black holes in Einstein's gravity, where $A =4\pi R^2$ represents
the area of the horizon. If we wish to include the quantum effects, we can assume the
following modification \cite{Sheykhi:2010wm, Sheykhi:2021fwh} 
\begin{equation}
 S=\frac{A}{4}+\mathcal{S}(A). 
 \end{equation}

The standard Bekenstein-Hawking result is reproduced when the second term vanishes. The entropy of the surface changes by one fundamental unit $\triangle S$ fixed by the discrete spectrum of
the area of the surface via the relation 
\begin{equation} 
d S=\frac{\partial S}{\partial A}d A=\left[\frac{1}{4}+\frac{\partial \mathcal{S}}{\partial A}\right]d A. 
 \end{equation}
Here we note that the energy of the surface $\Sigma$ is identified with the
relativistic rest mass $M$ of the source mass, $E=M$. On the surface $\Sigma$, we can relate the area of the surface to the number of bytes according to $ A=QN,$
where $Q$ is a fundamental
constant, and $N$ is the number of bytes. Let us assume that the temperature
on the surface is $T$, by means of the equipartition law
of energy \cite{Padmanabhan:2003pk}, we get the total energy on the surface via 
\begin{equation}
   E=\frac{1}{2}Nk_B T.
\end{equation}
We also need the force, which, according to this picture, it is the entropic force obtained from Eq. (1) where $\triangle S$ is one fundamental unit of entropy when
$|\triangle x|= \eta \lambda_m$, and the entropy gradient points
radially from the outside of the surface to the inside. Further taking $\triangle N=1$,and $\triangle A=Q$, we get 
\begin{equation}\label{F3}
F=-\frac{GMm}{R^2}\left(\frac{Q}{2\pi k_B
\eta}\right)\left[\frac{1}{4}+\frac{\partial \mathcal{S}}{\partial A}\right]_{A=4\pi R^2}.
\end{equation}
That is nothing but Newton's law of gravitation to the first
order provided we define $\eta=1/8\pi k_B$; we get $Q=1$. Thus we reach
\begin{equation}\label{F4}
F=-\frac{GMm}{R^2}\left[1+4\,\frac{\partial \mathcal{S}}{\partial A}\right]_{A=4\pi R^2}.
\end{equation}

Let us take the following non-singular Yukawa-type gravitational potential 
\begin{equation}
\phi(r)=-\frac{G M m}{\sqrt{r^2+l_0^2}}\left(1+\alpha\,e^{-\frac{r}{\lambda}}\right)|_{r=R},
\end{equation}
where $l_0$ is some small quantity of Planck length order $l_0 \sim 10^{-34} \; \rm {cm}$  and $\alpha>0$. The wavelength of
massive graviton reads \cite{Visser:1997hd} $\lambda=\frac{\hbar}{m_g c}>10^{20} \rm m $,
that leads to $m_g<10^{-64}$ kg for the graviton mass.  Actually, if we neglect the effects of $l_0$ and set it to zero, we get 
\begin{equation}
\phi(r)=-\frac{G M m}{r}\left(1+\alpha\,e^{-\frac{r}{\lambda}}\right)|_{r=R}.
\end{equation}
Such a potential can be obtained if we study the Einstein field equations $G_{\mu \nu}=8\pi G(T^{\rm matter}_{\mu \nu}+T_{\mu \nu}^{\rm field})$. At large distances, we can set $T^{\rm matter}_{\mu \nu}=0$, and focus on the effect of the gravitational field by linearizing the geometry $g_{\mu \nu}=\eta_{\mu \nu}+h_{\mu \nu}$. If we assume a nonzero mass for the graviton, in large distances with $(h\ll 1)$, one can obtain a Yukawa-like potential  $\phi(r) \sim ({\rm{const}}/r) \,\exp{(-r/\lambda)}$ (see for example  \cite{Visser:1997hd}). However, if we consider the total potential around the body with mass $M$, in that case, we need to take them for the potential the superposition of the standard potential $\Phi \sim -G M m/r$. We obtain an expression given by the last equation by choosing the constant in the Yukawa term, i.e., ${\rm {const}} \sim G M m \alpha$. It is interesting to note here that the Yukawa potential has been obtained in modified theories of gravity such as the $f(R)$ gravity in references \cite{Capozziello:2009vr, Benisty:2023qcv}.

Now by using the relation $F=-\nabla \phi(r)|_{r=R}$, we obtain the modified Newton's law at
of gravitation as
\begin{equation}\label{F5}
F=-\frac{G M m}{R^2}\left[1+\alpha\,\left(\frac{R+\lambda+l_0^2/R}{\lambda}\right)e^{-\frac{R}{\lambda}}\right]\left[1+\frac{l_0^2}{R^2}\right]^{-3/2}.
\end{equation}

In what follows, we shall neglect terms $\alpha l_0^2 \rightarrow 0$ as they are very small. For the total force, we obtain a modified Newton's law
\begin{equation}
F=- \frac{ G M m }{R^2} \left[1+\alpha\,\left(\frac{R+\lambda}{\lambda}\right)e^{-\frac{R} {\lambda}}\right]\left[1+\frac{l_0^2}{R^2}\right]^{-3/2}.
\end{equation}

Thus, with the  correction
in the entropy expression, we see that
\begin{equation}
1+\left(\frac{1}{2 \pi R}\right)\frac{d \mathcal{S}}{dR}=\left[1+\alpha\,\left(\frac{R+\lambda}{\lambda}\right)e^{-\frac{R}{\lambda}}\right].
\end{equation}
Solving for entropy  and setting $l_0 \rightarrow 0$ we obtain
\begin{equation}\label{entropy1}
S=\pi R^2-2 \pi  \alpha\left(R^2+3\lambda R+3\lambda^2 \right) e^{-\frac{R}{\lambda}} ,
\end{equation}
where the constant of integration is taken $S_0=\pi R^2$, for example, exponential corrections to entropy have been reported in \cite{Chatterjee:2020iuf}. 
Therefore, the Newtonian dynamics get modified due to the parameter
$\alpha$.  Therefore, we found that if the entropy is modified in the form of Eq. \eqref{entropy1}, Newton's law of gravity will be modified; in other words, we end up with a modified gravity law. The correction term in the expression for the entropy can be viewed as a volume law entanglement to the entropy due to the gravitons. That means that $\alpha$ results from the entanglement to the volume law entropy. 

\subsection{Flat rotational curves in galaxies}
One can easily check that the modified Newtonian dynamics can explain the flat rotation curves of galaxies. Using
the fact that $|F|=m\, v^2/r$, we can rewrite the circular speed of
an orbiting test object as
\begin{equation}
v^2 \simeq \frac{ G M }{r} \left[1+\alpha\,\left(\frac{r+\lambda}{\lambda}\right)e^{-\frac{r}{\lambda}}\right] \left[1+\frac{l_0^2}{r^2}\right]^{-3/2}.
\end{equation}
An interesting result is found for galactic scales when $r \leq\lambda$ and $e^{-\frac{r}{\lambda}} \rightarrow 1$, along with $l_0^2/r^2 \rightarrow 0$, we get
\begin{equation}
v^2 \simeq \frac{G M }{r}+\frac{G M (r+\lambda)\alpha}{r \lambda}.
\end{equation}

As a result, in the outer part of galaxies, $v^2
\approx \rm constant$, i.e., the second term is important and can be attributed to the dark matter effect, which implies that flat rotation curves of galaxies 
\begin{equation}
  v^2 \simeq \frac{G M (r+\lambda)\alpha}{r \lambda},  
\end{equation}
which is confirmed by observations and tends to be constant. We can further rewrite this term and re-obtain the MOND expression found by Milgrom,  $v^4=GM a_0$ \cite{Milgrom:1983ca, Milgrom:1983pn, Milgrom:1983zz}, and in our case, we have
\begin{equation}
    v^4 \simeq GM \left(  \frac{GM (r+\lambda)^2 \alpha^2}{r^2\lambda^2}\right),
\end{equation}
where  $M$ is the total baryonic mass of the galaxy, and it follows that
\begin{equation}
    a_0= \frac{GM (r+\lambda)^2\alpha^2}{r^2\lambda^2},
\end{equation}
gives the acceleration. That shows that one can reproduce MOND in this picture.

\subsection{Modified Friedmann equations}
It is exciting to see that we obtained the log corrections to
the entropy following quantum effects. That is the first
important results in the present work. Let us now extend our
discussion of the cosmological setup. Assuming the background
spacetime to be spatially homogeneous and isotropic, which is given
by the Friedmann-Robertson-Walker (FRW) metric
\begin{equation}
ds^2=-dt^2+a^2\left[\frac{dr^2}{1-kr^2}+r^2(d\theta^2+\sin^2\theta
d\phi^2)\right],
\end{equation}
where we can further use $R=a(t)r$, $x^0=t, x^1=r$, the two dimensional metric $ h_{\mu \nu}$. 
Here $k$ denotes the
curvature of space with $k = 0, 1, -1$ corresponding to flat, closed, and open universes, respectively. The dynamical apparent
horizon, a marginally trapped surface with vanishing expansion, is
determined by the relation
$h^{\mu
\nu}(\partial_{\mu}R)\,(\partial_{\nu}R)=0$. 
A simple calculation gives
the apparent horizon radius for the FRW universe
\begin{equation}
\label{radius}
 R=ar= {1}/{\sqrt{H^2+ {k}/{a^2}}}.
\end{equation}
For the matter source in the FRW universe, we shall assume a perfect
fluid described by the stress-energy tensor
\begin{equation}\label{T}
T_{\mu\nu}=(\rho+p)u_{\mu}u_{\nu}+pg_{\mu\nu}.
\end{equation}
On the other hand, the total mass $M = \rho V$ in the region
enclosed by the boundary $\mathcal S$ is no longer conserved, one can compute the change in the total mass using the pressure
$dM = -pdV$, and this leads to the continuity equation
\begin{equation}\label{Cont}
\dot{\rho}+3H(\rho+p)=0,
\end{equation}
with $H=\dot{a}/a$ being the Hubble parameter. Let us now derive the dynamical equation for Newtonian cosmology. Toward this goal, let us consider a compact spatial region $V$ with a compact boundary
$\mathcal S$, which is a sphere having radius $R= a(t)r$, where $r$ is a dimensionless quantity. Going back and combining the second law of Newton for the test
particle $m$ near the surface, with gravitational force (\ref{F5})
we obtain
\begin{equation}\label{F6}
m\ddot{a}r=-\frac{GMm}{R^2}\left[1+\alpha\,\left(\frac{R+\lambda}{\lambda}\right)e^{-\frac{R}{\lambda}}\right] \left[1+\frac{l_0^2}{R^2}\right]^{-3/2}.
\end{equation}
We also assume $\rho=M/V$ is the energy density of the matter
inside the the volume $V=\frac{4}{3} \pi a^3 r^3$. Thus, Eq.
(\ref{F6}) can be rewritten as
\begin{equation}\label{F7}
\frac{\ddot{a}}{a}=-\frac{4\pi G
}{3}\rho \left[1+\alpha\,\left(\frac{R+\lambda}{\lambda}\right)e^{-\frac{R}{\lambda}}\right] \left[1+\frac{l_0^2}{R^2}\right]^{-3/2}.
\end{equation}
this result represents the entropy-corrected dynamical equation for
Newtonian cosmology. To derive the modified Friedmann equations of FRW universe in general relativity, we can use the active gravitational mass $\mathcal M$ rather than the total mass $M$. It follows that, due to the
entropic corrections terms via the zero-point length, the active gravitational mass
$\mathcal M$ will be modified. Using Eq.
\eqref{F7} and replacing $M$ with $\mathcal M$, it follows 
\begin{equation}\label{M1}
\mathcal M =-\frac{\ddot{a}
a^2r^3}{G}\left[1+\alpha\,\left(\frac{R+\lambda}{\lambda}\right)e^{-\frac{R}{\lambda}}\right]^{-1} \left[1+\frac{l_0^2}{R^2}\right]^{3/2}.
\end{equation}
In addition, for the active gravitational mass, we can use the definition
\begin{equation}\label{Int}
\mathcal M =2
\int_V{dV\left(T_{\mu\nu}-\frac{1}{2}Tg_{\mu\nu}\right)u^{\mu}u^{\nu}}.
\end{equation}
From here, it is not difficult to show the following result
\begin{equation}\label{M2}
\mathcal M =(\rho+3p)\frac{4\pi a^3 r^3}{3}.
\end{equation}
Utilizing Eqs. (\ref{M1}) and (\ref{M2}) we  find
\begin{equation}\label{addot}
\frac{\ddot{a}}{a} =-\frac{4\pi G
}{3}(\rho+3p)\left[1+\alpha\,\left(\frac{R+\lambda}{\lambda}\right)e^{-\frac{R}{\lambda}}\right] \left[1-\frac{3\,l_0^2}{2\,R^2}\right].
\end{equation}
That is the modified acceleration equation for the dynamical
evolution of the  FRW universe. To simplify the work, since $l_0$ is a very small number, we can consider a series expansion around $x=1/\lambda$ via
\begin{eqnarray}
\left[1+\alpha\,\left(\frac{R+\lambda}{\lambda}\right)e^{-\frac{R}{\lambda}}\right]=1+\alpha-\frac{1}{2}\frac{\alpha R^2}{\lambda^2}+...
\end{eqnarray}
provided that $\alpha R^2/\lambda^2 \ll1$. This relation is justified since $\alpha \ll1$, and if we take for the graviton mass $m_g \sim 10^{-68}$ kg, it implies for $\lambda \sim 10^{26} $ m, which coincides with the radius of the observable Universe $R \sim 10^{26}$ m. Otherwise, in the region, $R \rightarrow \infty$, the exponential term vanishes, i.e. $e^{-R/\lambda} \rightarrow 0$, and we get the Friedman equations in standard General Relativity. It follows that the modified Friedmann equation for $\alpha R^2/\lambda^2 \ll1$ gives
\begin{equation}
\frac{\ddot{a}}{a}=- \left(\frac{4 \pi G }{3}\right)\sum_i \left(\rho_i+3p_i\right)\left[1+\alpha-\frac{1}{2}\frac{\alpha R^2}{\lambda^2}\right]\left[1-\frac{3\,l_0^2}{2\,R^2}\right], \label{Aaddot}
\end{equation}
where we have assumed several matter fluids with a constant equation of state parameters $\omega_i$ and continuity equations 
\begin{equation}\label{Cont_gen}
\dot{\rho}_i+3H(1+ \omega_i) \rho_i=0.
\end{equation}
Hence, we have the expression for densities $\ rho=\rho_{i 0} a^{-3 (1+\omega_i)}$. 
Eq. \eqref{Aaddot} becomes \begin{align}
\frac{\ddot{a}}{a} =& - \left(\frac{4 \pi G }{3}\right)\sum_i \left(1+3\omega_i\right) \rho_{i 0} a^{-3 (1+\omega_i)}\nonumber \\
\times &\left[1+\alpha-\frac{1}{2}\frac{\alpha R^2}{\lambda^2}\right]\left[1-\frac{3\,l_0^2}{2\,R^2}\right]. \label{2Aaddot}
\end{align}
Next, by multiplying $2\dot{a}a$ on both sides of Eq. \eqref{2Aaddot}, 
\begin{align}
2\dot{a}  \ddot{a} =& - \left(\frac{4 \pi G }{3}\right)\sum_i \left(1+3\omega_i\right) \rho_{i 0} a^{-3 (1+\omega_i)} 2\dot{a}a \nonumber \\
\times &\left[1+\alpha-\frac{1}{2}\frac{\alpha R^2}{\lambda^2}\right]\left[1-\frac{3\,l_0^2}{2\,R^2}\right], \\
d (\dot{a}^2+k)= & \frac{8\pi G}{3} \left[1+\alpha-\frac{1}{2}\frac{\alpha R^2}{\lambda^2}\right]\left[1-\frac{3\,l_0^2}{2\,R^2}\right] \nonumber \\
\times & d \left(\sum_i \rho_{i 0} a^{-1-3\omega_i)}\right), 
\end{align}
where $k$ is a constant of integration and physically characterizes the curvature of space.
Hence, with $R[a]= r a$ we have 
\begin{align}\label{Fried1}
 \dot{a}^2+k = &   \frac{8\pi G}{3} \int \left[1+\alpha-\frac{1}{2}\frac{\alpha R[a]^2}{\lambda^2}\right]\left[1-\frac{3\,l_0^2}{2\,R[a]^2}\right] \nonumber \\
\times & \frac{d \left(\sum_i \rho_{i 0} a^{-1-3\omega_i}\right)}{da} da,
\end{align}
 with $r$ nearly a constant.
 For several matter components with a constant equation of state, $\omega_i \notin\{-1, 1/3\}$, we have 
 \begin{align}
  & \frac{\dot{a}^2}{a^2}+\frac{k}{a^2}=  \frac{8\pi  G }{3} \left(\alpha  \left(\frac{3 l_0^2}{4\lambda ^2}+1\right)+1\right) \sum_i \rho_{i0} a^{-3 (1+\omega_{i})} \nonumber \\
  & -\frac{4 \pi  (\alpha +1) G l_0^2}{3 R^2} \sum_i\frac{ 3  \omega_{i}+1}{\omega_{i}+1}  \rho_{i0} 
a^{-3 (1+ \omega_{i})} \nonumber \\
   & +\frac{4 \pi  \alpha  G R^2}{3 \lambda ^2}  \sum_i \frac{ 1+ 3 \omega_{i}}{1-3  \omega_{i}}  \rho_{i0}  a^{-3 (1+\omega _{i})}.
 \end{align}
 Then, we obtain in leading order terms as $\frac{l_0^2}{\lambda ^2}\rightarrow 0$, 
\begin{align}
H^2+\frac{k}{a^2} = & \frac{8\pi G_{\rm eff}
}{3}\sum_i \rho_i -\frac{1}{R^2}  \sum_i \Gamma_1(\omega_i)\rho_i  \nonumber \\& +\frac{4 \pi G_{\rm eff}}{3}R^2 \sum_{i}\Gamma_2(\omega_i)\rho_i, \label{Fried01}
\end{align}
where 
\begin{equation}
G_{\rm eff}=G(1+\alpha),
\end{equation}
 along with the definitions
\begin{align}
\Gamma_1 (\omega_i ) & \equiv  \frac{4 \pi G_{\rm eff} l_0^2 }{ 3 }\left(\frac{1+3
\omega_i}{1+\omega_i}\right), \\
    \Gamma_2 (\omega_i )   & \equiv   \frac{\alpha\, (1+3\omega_i)}{  \lambda^2 (1+\alpha) (1-3\omega_i)},
\end{align}
provided $\omega_i \notin\{-1, 1/3\}$. Assuming only a  matter source, we obtain in leading order terms
\begin{equation}
H^2+\frac{k}{a^2} =\frac{8\pi G_{\rm eff}
}{3}\rho-\frac{\Gamma_1}{R^2}\rho+\frac{4 \pi G_{\rm eff}}{3}\rho\,\Gamma_2 R^2,
\end{equation}
where in the definitions for $\rho$, $\Gamma_1$, and $\Gamma_2$, we omitted the dependence of $\omega_i$. Using \eqref{Cont} and considering a flat universe ($k=0$), we have $R^2=1/H^2$, hence we can write 
\begin{equation}
H^2 \left(1+\Gamma_1 \rho\right)-\frac{4 \pi G_{\rm eff}}{3}\frac{\Gamma_2}{H^2}\rho=\frac{8\pi G_{\rm eff}
}{3}\rho.
\end{equation}
By expanding around $l_0$ and making use of $ \left(1+\Gamma_1 \rho\right)^{-1}\simeq \left(1-\Gamma_1 \rho\right)$ and neglecting the terms $\sim \mathcal{O}(l_0 \alpha^2/\lambda^2)$, we can finally write 
 \begin{equation}\label{imeq}
H^2-\frac{4 \pi G_{\rm eff}}{3} \frac{\Gamma_2}{ H^2} \rho  =\frac{8\pi G_{\rm eff} }{3}\rho\left(1-\Gamma_1 \rho \right).
\end{equation}

\subsection{Early time universe}
One can study two special cases for the Friedmann equation given by Eq. \eqref{imeq}. In the limit $\alpha \rightarrow 0$ [$\Gamma_2=0$], we get the quantum corrected Friedman's equations 
\begin{equation}\label{Fried2}
H^2=\frac{8\pi G }{3}\rho \left(1-\Gamma_1 \rho \right).
\end{equation}
This equation is essential to study the early Universe when the quantum effects are significant and was derived in \cite{Jusufi:2023ayv}. The phase space analyses were further studied in \cite{Millano:2023ahb}.

\subsection{Late time universe}
On the other hand, for the late Universe, we can neglect the quantum effects and set $l_0 \rightarrow 0$ [$\Gamma_1=0$]; we get 
\begin{equation}
H^2-\frac{4 \pi G_{\rm eff}}{3}\frac{\sum_i \Gamma_2(\omega_i)\,\rho_i}{H^{2}}=\frac{8\pi G_{\rm eff} }{3}\,\sum_i \rho_i,
\end{equation}
and after we use $\rho_{\rm crit}=\frac{3}{8 \pi G}H_0^2$,
we get two solutions 
\begin{align}\label{eq43}
   \frac{H^2}{H_0^2}&=\frac{(1+\alpha)}{2}\,\sum_i\Omega_i \notag \\
   &\pm  \frac{\sqrt{(\sum_i\Omega_i)^2 (1+\alpha)^2+2 \sum_{i} \frac{\Gamma_2(\omega_i) \Omega_i (1+\alpha)}{H_0^2}}}{2},
\end{align}
where  $\Omega_i=\Omega_{i0}(1+z)^{3(1+\omega_i)}, \, \Omega_{i0}=  8 \pi G \rho_{i0}/(3H_0^2)$ and the second sum under the radical runs for $\omega_i\neq 1/3$. 

A natural question arises: \textit{What is the physical interpretation of the term $
    \frac{ 2  \Gamma_2(\omega_i) \Omega_i (1+\alpha)}{H_0^2} $
in Eq. \eqref{eq43}}?  

As we shall show, this term precisely mimics the effect of dark matter in the $\Lambda$CDM model. Let us focus on the term proportional to $\Gamma_2 \Omega_i$ to obtain a similar effect as the cold dark matter. So it is natural to set $\omega_i=0$ (dust), i.e., having only baryonic matter in the form of dust with $\Omega_i=\Omega_B (1+z)^{3}$, and making use of $\Gamma_2$ we claim and define the following quantity [here we shall add the constant term $c$ to make the equation consistent]
\begin{equation}
   \frac{\Omega^2_{D}(1+\alpha)^2}{{(1+z)^3}}\equiv \frac{2 \Gamma_2 \Omega_i(1+\alpha)}{H_0^2}=\frac{2 \alpha\, c^2 \Omega_B\,{{(1+z)^3}} }{\lambda^2 H_0^2}.
\end{equation}
It follows that the dark matter can be viewed as an apparent effect, and it is obtained from  the baryonic matter
\begin{equation}\label{eqDM}
  \Omega_{D}= \frac{c}{\lambda H_0\,(1+\alpha)}\sqrt{2 \alpha \Omega_B} \,{(1+z)^{3} }.
\end{equation}
Using $\Omega_D=\rho_D/\rho_{\rm crit}$ and $\Omega_B=\rho_B/\rho_{\rm crit}$, one can further obtain the relation between the present densities for dark matter and baryonic matter in terms of the graviton mass
\begin{equation}
\rho_{D} = \frac{m_g c^2 }{2 \hbar (1+\alpha)  } \sqrt{\frac{3 \alpha \rho_B}{\pi G}}\,\Big|_{z=0}.
\end{equation}

Next, we assume a cosmological constant; therefore, we can expect a possible relation to the graviton mass. Let us define the following quantity
\begin{equation}
 \Omega_{\Lambda}= \frac{c^2}{\lambda^2 H^2_0}\frac{\alpha}{(1+\alpha)^2}.
\end{equation}
From this equation, we can further write the energy density of the cosmological constant contribution in terms of the graviton mass
\begin{equation}\label{gm}
\rho_{\Lambda} = \frac{3\,m_g^2 c^4\,\alpha}{8 \pi\, G\, \hbar^2\, (1+\alpha)^2 },
\end{equation}
which is indeed a constant quantity. In other words, dark matter and energy can be linked to the graviton mass. Finally, comparing the last equation with  $\rho_{\Lambda} = \frac{\Lambda c^2}{8 \pi\, G}$,
we can estimate the cosmological constant to be 
\begin{equation}
\Lambda = \frac{3\,m_g^2 c^2\,\alpha}{\hbar^2\, (1+\alpha)^2 }.
\end{equation}
In other words, a universe filled with gravitons with mass $m_g$ exactly produces the cosmological constant. Importantly, we found that the energy density is a constant quantity and can be viewed as a self-interaction effect between gravitons with mass $m_g$. 
Furthermore, we can combine the Eq. \eqref{eqDM} with the last equation, and we can relate baryonic matter, apparent dark source, and the cosmological constant
\begin{equation}\label{DMCC}
\Omega_D= \sqrt{2\, \Omega_B  \Omega_{\Lambda}}{(1+z)^3}.
\end{equation}
The last equation gives another interpretation for dark matter: dark matter can be viewed as an apparent effect due to the interaction of baryonic matter and the cosmological constant. By including the cosmological constant in Eq. \eqref{eq43}, we now have 
\begin{align}
   {H^2(z)}/{H_0^2}&=\frac{(1+\alpha)}{2} \Big(\Omega_B (1+z)^{3}+\Omega_{\Lambda}\notag  \\ 
   &+ \sqrt{(\Omega_B (1+z)^{3}+\Omega_{\Lambda})^2+\frac{\Omega^2_D}{(1+z)^3}}\,\Big). 
\end{align}
It can be seen that effectively the $\Lambda$CDM can be obtained for example if we consider an expansion around $\Omega^2_{DM}/(1+z)^3$; then in leading-order terms that yields
\begin{eqnarray}
    {H^2(z)}/{H_0^2}&=&(1+\alpha) \Big(\tilde{\Omega}_B (1+z)^{3}+\Omega_{\Lambda}\Big).
\end{eqnarray}
where $\tilde{\Omega}_{B,0}=\Omega_{B,0}(1+\frac{1}{2 \left(1+\Omega_{B,0}(1+z)^{3}/\Omega_{\Lambda,0}\right)}),$ which shows that the dark matter contribution is absorbed in the first $\tilde{\Omega}_{B,0}$. Alternatively one may define a total quantity $\Omega^2$ viewed as the root-mean-square density energy to obtain a similar effective equation. The last equation shows that the $\Lambda CDM$ model is effectively obtained from our model. Again, this equation shows that in our model, we only have baryonic matter as a real type of matter. The dark matter is only an apparent effect and results from the coupling between baryonic matter via the Yukawa gravitational potential having a graviton with nonzero mass $m_g$. On the other hand, the cosmological constant term is also related to the graviton mass and $\alpha$. It can be viewed as a self-interaction effect between gravitons. In the limit $\alpha \to 0$, we have $\Omega_D=\Omega_{\Lambda}=0$, and the standard $\Lambda$CDM model of general relativity with only baryonic matter is obtained. Let us see an important difference from the standard $\Lambda$CDM model. To obtain the $\Lambda$CDM-like model for the total matter, we need to take 
\begin{eqnarray}\notag
   (1+\alpha) \tilde{\Omega}_B (1+z)^{3} &\to & \Omega^{\Lambda CDM}_B(1+z)^3+\Omega_D^{\Lambda CDM},\\\notag
   (1+\alpha) \Omega_{\Lambda} &\to & \Omega_{\Lambda}^{\Lambda CDM},\\
  (1+\alpha) \Omega_{D} &\to & \Omega_D^{\Lambda CDM}.
\end{eqnarray}
Hence, to get the $\Lambda$CDM-like model, we can write
 \begin{small}
 \begin{equation}
    \frac{H^2(z)}{H_0^2}=\left(\Omega^{\Lambda CDM}_B + \sqrt{2 \, \Omega_\Lambda^{\Lambda CDM} \, \Omega_B^{\Lambda CDM}} \right)(1+z)^{3} +\Omega_{\Lambda}^{\Lambda CDM},
\end{equation}
\end{small}
or alternatively,
\begin{equation}
    \frac{H^2(z)}{H_0^2}= \Omega^{\Lambda CDM}_B (1+z)^{3} +\Omega_D^{\Lambda CDM}+\Omega_{\Lambda}^{\Lambda CDM},
\end{equation}
with 
\begin{eqnarray}
\Omega_{\Lambda}^{\Lambda CDM}&=&\frac{c^2 \alpha}{\lambda^2 H_0^2  (1+\alpha)  },\\
\Omega_D^{\Lambda CDM} &=&\sqrt{2 \Omega_{B}^{\Lambda CDM} \Omega_{\Lambda}^{\Lambda CDM} },
\end{eqnarray}
along with $H_0=H_0^{\Lambda CDM}$, in terms of these definitions.
In the standard $\Lambda$CDM model, we usually add the cold dark matter and dark energy terms by hand. The main argument for doing this is the assumption that a dark matter particle exists and, conversely, the existence of vacuum energy. Of course, adding such terms by hand is problematic since there has yet to be evidence of dark matter particles. Nevertheless, as we can see in our model, there is no dark matter particle at all; the dark matter term naturally appears due to modifying the law of gravity. In addition, the cosmological constant parameter, as we argued, also can be expressed in terms of parameters $\lambda$ and $\alpha$. 

\subsection{Correspondence with Verlinde's emergent gravity and non-local gravity}
 One of the most important findings in this paper is the relation between the dark matter and the baryonic matter given by Eq. \eqref{eqDM}. Here we shall point out that this relation is very similar to the relation found by Verlinde in the emergent gravity (EMG) scenario where for constant densities, it was found  \cite{Verlinde:2016toy}
\begin{equation}\label{dmv}
    \Omega^{EMG}_{D}=2 \sqrt{\Omega^{\Lambda CDM}_B/3}.
\end{equation}
That means that there exists a precise correspondence with our model when $\Omega_D^{\Lambda CDM} \to  \Omega^{EMG}_D$,  hence we get
\begin{equation}
  \frac{c}{\lambda H_0  (1+\alpha)}\sqrt{2\, \alpha}\,\Big|_{z=0} \longrightarrow  {2}/{\sqrt{3}} \simeq 1.15.
\end{equation}
Further if we combine  Eq. \eqref{DMCC} and \eqref{dmv} for constant densities we get $\Omega^{\Lambda CDM}_{\Lambda} \to {2}/{3}$. In the general case, as we have shown in \cite{Jusufi:2023ayv}, we can generalize Verlinde's relation by writing 
\begin{equation}
 \sqrt{2\,\Omega^{\Lambda CDM}_B\Omega^{\Lambda CDM}_\Lambda}\Big|_{z=0} \longrightarrow  \sqrt{\frac{2\,\Omega^{\Lambda CDM}_B\,(2\,\alpha)}{3}},
\end{equation}
which implies
\begin{eqnarray}
    \Omega_{\Lambda}^{\Lambda CDM}=\frac{2}{3}\alpha.
\end{eqnarray}
Using the last equation and $\Omega_{\Lambda}^{\Lambda CDM}=\rho_{\Lambda}/\rho_{\rm crit}$, one can obtain the cosmological constant in terms of $\alpha$, 
\begin{eqnarray}
    \Lambda=\frac{2 \alpha H_0^2}{c^2}.
\end{eqnarray} 
Further, by making use of the relation between the Hubble constant and the  Hubble scale via $H_0=c/L$ \cite{Verlinde:2016toy}, we can obtain 
\begin{eqnarray}
    \Lambda=\frac{2 \alpha}{L^2}.
\end{eqnarray} 
In this way, we obtain a holographic dark energy expression for the cosmological constant 
\begin{eqnarray}
    \rho_{\Lambda}=\frac{\alpha \,c^2}{4 \pi G L^2}.
\end{eqnarray}

In addition, in a recent work \cite{Jusufi:2023ayv}, it was argued that Verlinde's emergent gravity could be viewed as a non-local gravity theory. Specifically, apparent dark matter may be a consequence of the emergent nature of gravity and caused by an
elastic response due to the volume law contribution to the
entanglement entropy in our universe \cite{Verlinde:2016toy}. Verlinde's theory also has a cosmological constant due to the vacuum energy. In our case, the energy density of the cosmological constant is related to the graviton mass, as we can see from Eq. \eqref{gm}. This non-local manifestation appears naturally in our model as well. That has to do with the long-range force interaction, which makes the theory non-local. In such a case, the field equations reduce to
\cite{Hehl:2008eu,Blome:2010xn,Chicone:2015sda}
\begin{equation}\label{Ba2}
  G_{\mu\nu}(x)+\int{\cal K}(x,y)G_{\mu\nu}(y)d^4y=\kappa T_{\mu\nu}\,,
\end{equation}
in which $G_{\mu\nu}$ is the linear
Einstein tensor and $T_{\mu\nu}$ is the
energy-momentum tensor of matter with a
nonlocal Poisson equation  \cite{Hehl:2008eu,Blome:2010xn,Chicone:2015sda}
\begin{equation}\label{Newton}
\nabla^2\Phi=4\pi G\left[\rho(t,\mathbf{x})+\rho_{\rm D}(t,\mathbf{x})\right]\,,
\end{equation}
where the ``density of DM'' $\rho_{\rm D}$ is given by
\begin{equation}\label{dark}
\rho_{\rm D}(t,\mathbf{x})=\int q(\mathbf{x}-\mathbf{y})
\rho(t,\mathbf{y})d^3y\,.
\end{equation}
In the last equation, the quantity $q$ is expected to be a universal function that is
independent of the nature of the source. It is not difficult to show that dark matter emerges from the non-local aspect of the gravitational interaction. We can further use these relations to explain the
the circular motion of stars and the observed flat rotation
curves in galaxies by making use of the kernel \cite{Hehl:2008eu, Blome:2010xn, Chicone:2015sda}
\begin{equation}\label{qernel}
q(\mathbf{x}-\mathbf{y})=\frac{1 }{4\pi \beta}
\frac{1}{|\mathbf{x}-\mathbf{y}|^2}\,,
\end{equation}
where $\beta=GM/v^2$ reflects the universality
of the non-local kernel meaning that we obtain a constant term
 $M\propto v^2$. In particular, the potential energy per unit
test mass is given by
\begin{equation}\label{nonlocal}
\phi=-\frac{G M }{r} \left(1- \frac{r}{\beta}\, \ln\left(\frac{r}{\beta}\right)\right).
\end{equation}

If we perform a series expansion of the Yukawa potential around $l_0$, we get 
\begin{equation}
\phi=-\frac{G M}{r}\left(1+\alpha\,e^{-\frac{r}{\lambda}}\right)+...
\end{equation}
To see the correspondence, we can write the approximation
\begin{equation}
    e^{-\ln\left[\left(\frac{r}{\beta}\right)^{\frac{r}{\beta}}\right]}\simeq 1- \frac{r}{\beta} \ln \left(\frac{r}{\beta}\right)+...
\end{equation}
In this sense, the potential \eqref{nonlocal} approximates a more general expression for the potential: the Yukawa potential. Using Eq. (15) and $\beta=GM/v^2$, we also can see that there exists a relation between the parameters $\beta$ and $\lambda$ as follows $\beta=r \lambda/ [(r+\lambda) \alpha]$, explaining the origin of the large-scale gravity modification in non-local gravity related to the parameter $\lambda$. The appearance of dark matter as an emergent effect or modification of gravity can also be seen in the scalar-tensor gravity theory picture \cite{Moffat:2020ffz}. In particular, as pointed out in \cite{Jusufi:2023ayv}, one can investigate the gravitational interaction at various scales ranging from clusters of galaxies to cosmology. The main point is that as soon as we  ``localize" the non-linear gravity, corrections immediately arise. Finally, we have the effect of a scalar field ruling the gravitational interaction \cite{Moffat:2020ffz}.

\subsection{Observational constrains}

The companion paper \cite{Gonzalez:2023rsd} is entirely devoted to observational constraints, and there were compared $m_g$ values with the ones in the literature (see Table 3 and Figure 5 in \cite{Gonzalez:2023rsd}). Having in mind that $\lambda$ should be a huge parameter, in the present paper, we guessed $\lambda$ to be of $\rm {Gpc}$ order, namely $\lambda \sim 10^{3}\;\rm {Mpc}$, which seems to work very well. We then use the value for $H_0$ to extract the range for $\alpha$.

Other studies have estimated the graviton mass differently. For instance, A. S. Goldhaber \textit{et al.} (1974) \cite{Goldhaber:1974wg}, using data from Galaxy Clusters, obtained $\lambda>0.324\;\rm {Mpc}$ and $m_g<10^{-38}\;\rm {GeV}$. C. Talmadge \textit{et al.} (1998) \cite{Talmadge:1988qz}, using data from Solar System measurements, obtained $\lambda >9.074\times 10^{-8}\;\rm {Mpc}$ and $m_g<7.2\times 10^{-32} \;\rm {GeV}$. Moreover, S. R. Choudhury \textit{et al.} (2004) \cite{Choudhury:2002pu}, using Weak Lensing measurements, obtains $\lambda>97.22\;\rm {Mpc}$ and $m_g<6\times 10^{-41}\;\rm {GeV}$. E. Berti \textit{et al.} (2011) \cite{Berti:2011jz} observed multiple inspiralling black holes to determine the bounds that space-based detectors could place on the graviton Compton wavelength and graviton mass, leading to $\lambda> 9.722\times 10^{-4}\;\rm {Mpc}$ and  $m_g<4 \times 10^{-35} \;\rm {GeV}$. 
L. Shao \textit{et al.} (2020) \cite{Shao:2020fka} shows $\lambda\gtrsim  0.2268\;\rm {Mpc}$  and  $m_g \lesssim 2 \times 10^{-37} \;\rm {GeV}$.    D. Benisty (2022) \cite{Benisty:2022txp} shows a Keplerian-type parametrization as a solution of Yukawa-type potential accurate equations of motion for two non-spinning compact objects moving in an eccentric orbit. A bond from the solar system is presented leading to Solar system  $\alpha=(3.863 \pm 0.373)  10^{-3}$, $m_g=(1.406 \pm 0.801) \times 10^{-30}\;\rm {GeV}$   (Solar system)    and $\alpha=(4.351 \pm 0.2713) 10^{-7}$, $m_g=(4.384 \pm 2.146) \times 10^{-28}\;\rm {GeV}$ (Solar system + Cassini). D. Benisty \textit{et al.} (2023) \cite{Benisty:2023ofi} obtain a value of $\alpha<0.581$ and $m_{g}<5.095\times 10^{-35}$ with Local Group estimations. Finally, E. Gonzalez \textit{et al.} (2023)  using SNe Ia+OHD data found $\alpha= 0.416_{-0.326}^{+1.137}$, $\lambda= 2693_{-1262}^{+1191}\;\rm {Mpc}$ and  $m_g=\left(2.374_{-0.728}^{+2.095}\right)\times 10^{-42}\;\rm {GeV}$, that is consistent with our preliminary results.

Table \ref{tab:Comparison-Yukawa_parameters} complements other significant findings related to Yukawa constraints from different systems \cite{Goldhaber:1974wg, Talmadge:1988qz, Choudhury:2002pu, Berti:2011jz, Shao:2020fka, Benisty:2022txp, Benisty:2023ofi, Gonzalez:2023rsd}. See also the summary in Table I of the review \cite{deRham:2016nuf}. 
These are current and projected bounds on the graviton mass and its reduced Compton wavelength. The masses reported are upper bounds and the reduced Compton wavelengths' lower bounds.

\begin{table*}[ht!]
    \centering
        \begin{tabularx}{\textwidth}{YYYYY}
        \hline\hline
        \multirow{2}{*}{Data} &\multicolumn{3}{c}{Constraint} & \multirow{2}{*}{Reference} \\
        \cline{2-4}
         & $\alpha$ & $\lambda\;\rm {[Mpc]}$ & $m_{g}\;\rm {[GeV]}$ & \\
        \hline \\
        Galaxy Clusters& $\cdots$ & $>0.324$ & $<10^{-38}$ & A. S. Goldhaber \textit{et al.} (1974) \cite{Goldhaber:1974wg} \\
        Solar System& $\cdots$ & $>9.074\times 10^{-8}$ & $<7.2\times 10^{-32}$ & C. Talmadge \textit{et al.} (1998) \cite{Talmadge:1988qz} \\
        Weak Lensing& $\cdots$ & $>97.22$ & $<6\times 10^{-41}$ & S. R. Choudhury \textit{et al.} (2004) \cite{Choudhury:2002pu} \\
        BH  binaries& $\cdots$ & $> 9.722\times 10^{-4}$ & $<4 \times 10^{-35}$ & E. Berti \textit{et al.} (2011) \cite{Berti:2011jz}\\
        Binary Pulsars&     $\cdots$ &  $\gtrsim  0.227$  &   $\lesssim 2 \times 10^{-37}$ & L. Shao \textit{et al.} (2020) \cite{Shao:2020fka}\\
        Solar system&  $(3.863 \pm 0.373)  10^{-3}$        &    $\cdots$     &        $(1.406 \pm 0.801) \times 10^{-30}$ & D. Benisty (2022)  \cite{Benisty:2022txp}\\ 
        Solar system +Cassini    &  $(4.351 \pm 0.2713) 10^{-7}$         &    $\cdots$      &       $(4.384 \pm 2.146) \times 10^{-28}$ & D. Benisty (2022)  \cite{Benisty:2022txp}\\   
        Local Group & $<0.581$ & $\cdots$ & $\left(3.265 _{-1 . 830}^{+1 . 830} \right)\times 10^{-35}$ & D. Benisty \textit{et al.} (2023) \cite{Benisty:2023ofi} \\
        SNe Ia+OHD & $0.416_{-0.326}^{+1.137}$ & $2693_{-1262}^{+1191}$ & $\left(2.374_{-0.728}^{+2.095}\right)\times 10^{-42}$ & E. Gonzalez \textit{et al.} (2023)  \cite{Gonzalez:2023rsd} \\ 
        Educated Guess           &         $\alpha \simeq 0.04$                 &      $ \simeq  10^3$                 &  $6.418\times 10^{-42}$ & This paper  \\
       \hline\hline
    \end{tabularx}
    \caption{Estimations of the coupling constant $\alpha$, the wavelength parameter ($\lambda$), and graviton mass ($m_g$) for the Yukawa potential. }
    \label{tab:Comparison-Yukawa_parameters}
\end{table*}

\section{Phase space analyses}
\label{Sect:III}

In this section, we aim to study the phase space analyses of the modified Friedmann equation obtained in the last section and given by \eqref{imeq}. 
When $\Gamma_2=0$, we obtain the early universe evolution in which the quantum effects encoded in the term $\Gamma_1$ play an important role. The phase space analyses for such a universe were recently studied in \cite{Millano:2023ahb}. Raychaudhuri equation becomes
\begin{align}
    \dot{H} & =  -\frac{6 \pi  G_{\rm eff} (\omega+1) H^2 \rho 
   \left( \Gamma_{2}+2H^2 (1-2  \Gamma_{1} \rho)\right)}{4 \pi  \Gamma_{2} G_{\rm eff} \rho+ 3H^4}, \label{eq73}
\end{align}
together with the equation of motion \eqref{Cont}, 
with first integral \eqref{imeq}.

\subsection{Case $\Gamma_2 <0, \Gamma_1<0$}
 
$\Gamma_2 <0, \Gamma_1<0$ implies 
$-1<\omega<-\frac{1}{3}$. Then, we define 
\begin{align}
    x &= \frac{\sqrt{3} H}{\sqrt{8 \pi G_{\rm eff}  \rho\left(1 + |\Gamma_1| \rho \right)}}, \;
    y = \frac{\sqrt{|\Gamma_2|}}{H \sqrt{2 \left(1 + |\Gamma_1| \rho \right)}}, \\
    \Omega & = \frac{8 \pi  G_{\rm eff} \rho}{3 H^2}, \label{Omega}
\end{align}
and for $\omega>-1$ we define the new derivative $\tau= (a/a_0)^{3(1+\omega)}$. 
Using the relation $x^2+y^2=1$, we obtain de decoupled system 
\begin{align}
& \frac{dy}{d\tau}= \frac{(y-1) y (y+1) \left(2 y^2 \Omega -2 \Omega +3\right)}{2 \left(2 y^2-1\right)}, \label{DS1}\\
& \frac{d\Omega}{d\tau}= -\frac{\Omega  \left(y^2 \Omega +y^2-\Omega +1\right)}{2 y^2-1}. \label{DS2}
\end{align}
Defined in the compact space $(y,\Omega)\in [-1,1]\times [0,1]$. 

\begin{figure}[t!]
    \centering
    \includegraphics[scale=0.9]{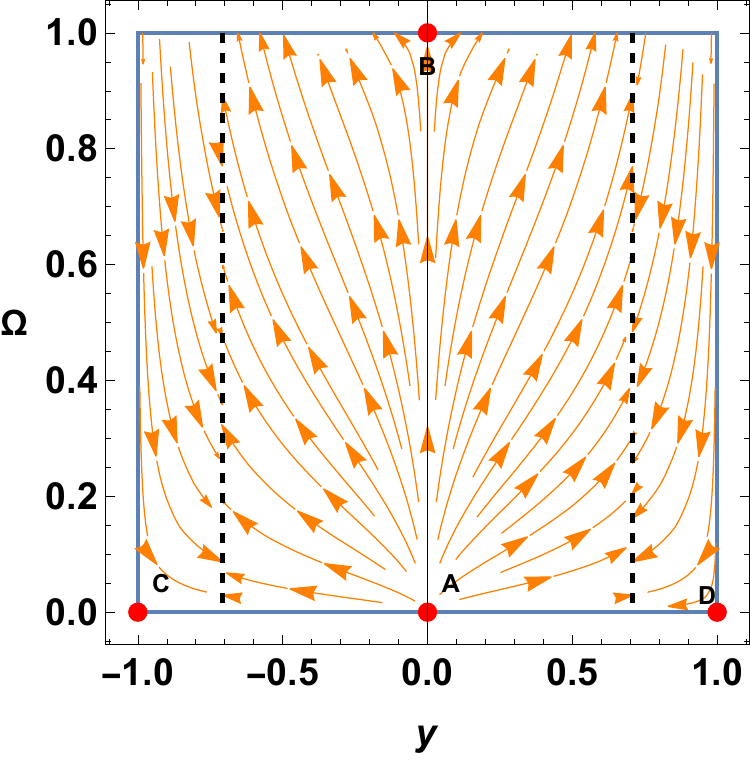}
    \caption{Phase plot of system \eqref{DS1}-\eqref{DS2}.}
    \label{fig:case A}
\end{figure}

Tab. \ref{tab:1} shows the equilibrium points for the dynamical system \eqref{DS1}-\eqref{DS2}, and  Fig. \ref{fig:case A} depicts the phase plot.

\begin{table}[h!]
    \centering
  \setlength{\tabcolsep}{4pt}
\begin{tabular}{cccccc}\hline 
Label & $y$ & $\Omega$ & $\lambda_1$ & $\lambda_2$ & Stability\\\hline 
$A$ & $0$ & $0$ & $\frac{3}{2}$ & $1$ & source \\\hline 
$B$ & $0$ & $1$ & $-1$ & $\frac{1}{2}$ & saddle \\ \hline 
$C$ & $-1$ & $0$ & $3$ & $-2$ & saddle \\ \hline 
$D$ & $1$ & $0$ & $3$ & $-2$ & saddle\\\hline 
$E_\pm$ & $\pm  {\sqrt{2}}/{2}$ & arbitrary  & $-$ & $-$ & sink\\\hline
\end{tabular}
    \caption{Equilibrium points for the dynamical system \eqref{DS1}-\eqref{DS2}.}
    \label{tab:1}
\end{table}
 
The point $A$ corresponds asymptotically to the quantum corrected Friedman's equations \eqref{Fried2} such that ${\Gamma_2} \rho /{ H^2} \ll 1$. 
Because the point is the source, the quadratic density modification dominates in the early Universe. The saddle point $B$ corresponds asymptotically to $H^2= \frac{8 \pi G_{\rm eff} }{3} \rho$.
That is the usual dark matter-dominated FLRW universe. 
Saddle points $C$ and $D$ corresponds asymptotically to 
$2 H^2 = |\Gamma_2|/\left(1 + |\Gamma_1| \rho \right)$. Finally, there exist attractor lines $E_\pm$ with $x^2=y^2=1/2$, $
\frac{|\Gamma_2|}{\rho\left(1 + |\Gamma_1| \rho \right)^2}=\frac{4 \pi G_{\rm eff}  }{3}, H^2=\frac{|\Gamma_2|}{ \left(1 + |\Gamma_1| \rho \right)}$, representing de Sitter solutions.

 \subsection{Case $\Gamma_2 >0, \Gamma_1>0$}
$\Gamma_2 >0, \Gamma_1>0$ implies $ -\frac{1}{3}<\omega<\frac{1}{3}$. Then, we define 
\begin{align}
    x &=\frac{\sqrt{3} H^2}{\sqrt{4 \pi  G_{\rm eff}  \rho \left(\Gamma_{2}+2 H^2\right)}}, \quad
    y = \frac{\sqrt{2 \Gamma_{1}} H \sqrt{\rho}}{\sqrt{\Gamma_{2}+2 H^2}},
\end{align}
and $\Omega$ through \eqref{Omega}.
Using the relation $x^2+y^2=1$, we obtain de decoupled system 
\begin{align}
& \frac{dy}{d\tau}=\frac{y \left(y^2-1\right) \left(2 \left(y^2-1\right) \Omega +3\right)}{2 y^2 (\Omega -1)-2
   \Omega +4},\label{2DS1}
\\
& \frac{d\Omega}{d\tau}=-\frac{\Omega  \left(y^2 (\Omega +1)-\Omega +1\right)}{y^2 (\Omega -1)-\Omega
   +2}, \label{2DS2}
\end{align}
defined in the compact space $(y,\Omega)\in [-1,1]\times [0,1]$.

\begin{figure}[t!]
    \centering
\includegraphics[scale=0.9]{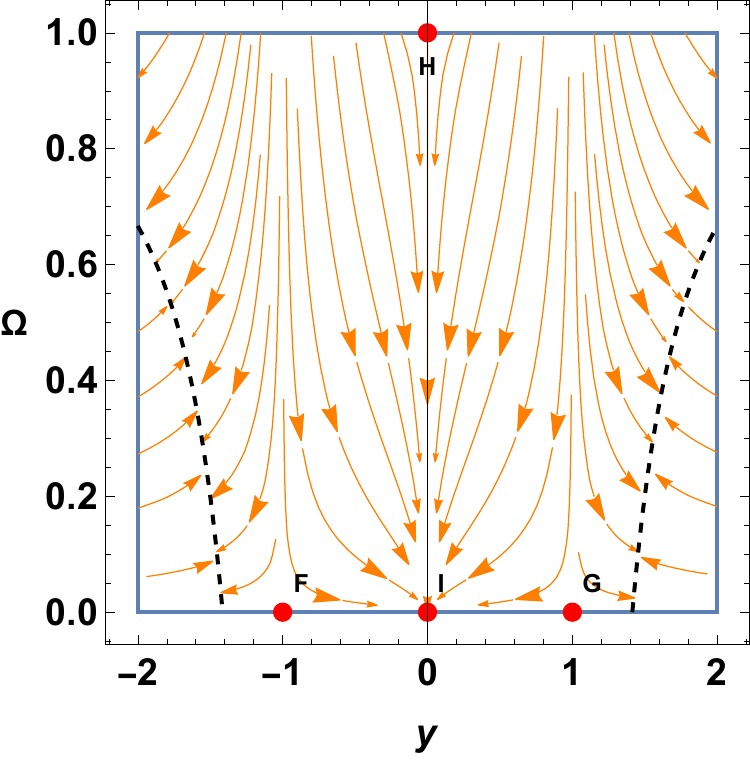}
    \caption{Phase plot of system \eqref{2DS1}-\eqref{2DS2}. }
    \label{fig:caseB}
\end{figure}

Tab.  \ref{tab:2} shows the equilibrium points for the dynamical system \eqref{2DS1}-\eqref{2DS2} and 
 Fig. \ref{fig:caseB} depicts the phase plot.

\begin{table}[h!]
    \centering
  \setlength{\tabcolsep}{4pt}
\begin{tabular}{cccccc}\hline 
Label & $y$ & $\Omega$ & $\lambda_1$ & $\lambda_2$ & Stability\\\hline
$F$ & $-1$ &$ 0 $& $3$ & $-2$ & saddle \\\hline
$G$ & $1$ & $0$ & $3$ & $-2$ & saddle \\\hline
$H$ & $0$ & $1$ & $1$ & $-\frac{1}{2}$ & saddle \\\hline
$I$ & $0$ & $0$ & $-\frac{3}{4}$ & $-\frac{1}{2}$ & sink \\\hline
$J$ & $y_c$ & $(2-y_c^2)/(1-y_c^2)$  & $-$ & $-$ & sink\\\hline
\end{tabular}
    \caption{Equilibrium points for the dynamical system \eqref{2DS1}-\eqref{2DS2}.}
    \label{tab:2}
\end{table}

The equilibrium points $F$ and $G$ are saddles. They correspond to $\rho = \frac{1}{\Gamma_1}+\frac{\Gamma_2}{2 \Gamma_1 H^2}$ and $\rho\ll  H^2$, e.g., $\rho \rightarrow \frac{1}{\Gamma_1}, H\rightarrow \infty$. Point 
$H$ corresponding to $\Gamma_{2}\ll 2 H^2$ and $H^2=\frac{8\pi  G_{\rm eff} }{3} \rho$  is a matter-dominated solution and it is a saddle. Point 
$I$ corresponds to $\rho= 3 H^2/(4 G_{\rm eff}  \pi (2  + \Gamma_2/H^2))$, and $\Omega=0$ is a sink. 
 
 Finally, the system admits the attractor line of equilibrium points $J: (y,\Omega)=\left(y_c,\frac{2-y_c^2}{1-y_c^2}\right)$, representing de Sitter solutions. 
  
 \subsection{Case $\Gamma_2 <0, \Gamma_1>0$}
$\Gamma_2 <0, \Gamma_1>0$ implies 
$w>\frac{1}{3}$.  Then, we define
\begin{align}
& x=  \sqrt{\Gamma_{1} \rho}, \quad y=  \sqrt{|\Gamma_{2}|}/(\sqrt{2} H), 
\end{align}
and $\Omega$ through \eqref{Omega}.
Using the relation  
 \begin{equation}
   x^2+y^2+ {1}/{ \Omega  }=1,
 \end{equation}
 we obtained de decoupled system
\begin{align}
& \frac{dy}{d\tau}=-\frac{y \left(\left(y^2-1\right) \Omega +2\right)}{2 y^2 \Omega -2}, \label{3DS1}\\
& \frac{d\Omega}{d\tau}=-\frac{\Omega \left(\left(2 y^2-1\right) \Omega +1\right)}{y^2 \Omega -1}. \label{3DS2}
   \end{align}

\begin{figure}[h!]
    \centering
\includegraphics[scale=0.9]{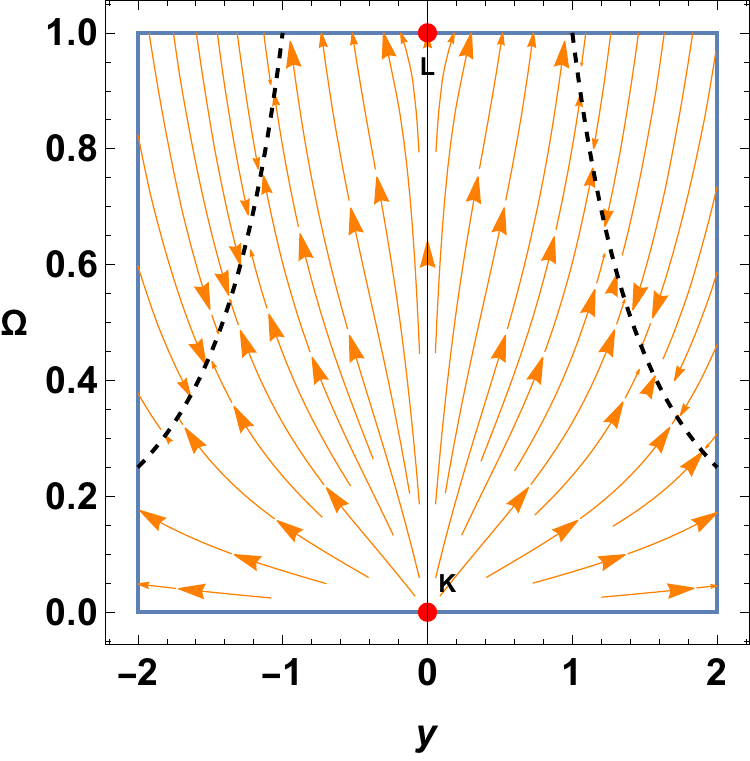}
    \caption{Phase plot of system \eqref{3DS1}-\eqref{3DS2}. }
    \label{fig:CaseC}
\end{figure}

Tab. \ref{tab:3} shows the equilibrium points for the dynamical system \eqref{3DS1}-\eqref{3DS2} and 
 Fig.  \ref{fig:CaseC} depicts a phase plot. 

\begin{table}[h!]
    \centering
  \setlength{\tabcolsep}{4pt}
\begin{tabular}{cccccc}\hline
Label & $y$ & $\Omega$ & $\lambda_1$ & $\lambda_2$ & Stability\\\hline
$K$ & $0$ & $0$ & $1$ & $1$ & source\\\hline
$L$ & $0$ & $1$ & $-1$ & $\frac{1}{2}$ & saddle \\\hline
$M$ & $y_c$  & $1/y_c^2$ & $-$ & $-$ & sink\\\hline
\end{tabular}
    \caption{Equilibrium points for the dynamical system \eqref{3DS1}-\eqref{3DS2}.}
    \label{tab:3}
\end{table}

The point $K$ is a source corresponding to $\rho, H\rightarrow \infty$. The point $L$ is a matter-dominated solution with $\rho\rightarrow 0$ and $H\rightarrow \infty$. Finally, the system admits the line of equilibrium points $M: (y,\Omega)=\left(y_c,\frac{1}{y_c^2}\right)$, which is a sink corresponding to de Sitter solutions. 

\subsection{Comparison with $\Lambda$CDM}

To compare with the results from $\Lambda$CDM, we define the quantities 
\begin{equation}
  E= {H}/{H_0}, \quad    \Omega_X  =8 \pi G \rho/(3 H^2),
 \end{equation}
where  $\Omega_X$ is identified with the dimensionless matter density, with energy density $\rho_i= \rho_X$ of an unknown fluid with a constant equation of state $\omega$, i.e., $p_i= \omega\rho_X$, and $E$ is the dimensionless Hubble parameter,
 and we defined the  constants
 \begin{equation}
 (\bar{\lambda}, \; \bar{l})= H_0 (\lambda/c, \; l_0/c), 
 \end{equation} and using the equations \eqref{Cont_gen} and \eqref{eq73} we 
 construct the system 
\begin{widetext}
 \begin{align}
  E'(z) & =\frac{3 E \Omega_X \left((\omega +1) \left(\alpha  \left(2 c^2 E^2 (3 \omega -1) \bar{\lambda }^2-3 \omega -1\right)+2 c^2 E^2 (3 \omega -1)
   \bar{\lambda }^2\right)-2 (\alpha +1)^2 c^4 E^4 \left(9 \omega ^2-1\right) \Omega_X \bar{\lambda }^2 \bar{l}^2\right)}{(z+1) \left(4 c^2 E^2 (3 \omega -1)
   \bar{\lambda }^2-2 \alpha  (3 \omega +1) \Omega_X\right)}, \label{eqE}\\
     \Omega_X'(z) & =\frac{6 c^2 E^2 \Omega_X \bar{\lambda }^2 \left((\alpha +1)^2 c^2 E^2 \left(9 \omega
   ^2-1\right) \Omega_X^2 \bar{l}^2-(\alpha +1) (\omega +1) (3 \omega -1) \Omega_X+3 \omega ^2+2 \omega -1\right)}{(z+1) \left(2 c^2 E^2 (3 \omega -1) \bar{\lambda
   }^2-\alpha  (3 \omega +1) \Omega_X\right)}. \label{eqOmegaX}
 \end{align}  
\end{widetext}
For simplicity, we choose $\omega=0$, such that $\Omega_X= \Omega_m$ corresponds to the dimensionless energy density of a pressureless matter. 
The system \eqref{eqE}--\eqref{eqOmegaX} reduces to 
\begin{small}
\begin{align}
   E^{\prime} & =\frac{3 E \Omega _m \left(2 (\alpha +1) c^2 E^2 \bar{\lambda }^2 \left(1-(\alpha +1) c^2 E^2 \bar{l}^2 \Omega _m\right)+\alpha \right)}{2 (z+1) \left(2 c^2
   E^2 \bar{\lambda }^2+\alpha  \Omega _m\right)}, \label{FeqE}\\
    \Omega_m^{\prime}& =\frac{6 c^2 E^2 \bar{\lambda }^2 \Omega _m \left((\alpha +1) \Omega _m \left((\alpha +1) c^2 E^2 \bar{l}^2 \Omega
   _m-1\right)+1\right)}{(z+1) \left(2 c^2 E^2 \bar{\lambda }^2+\alpha  \Omega _m\right)},\label{FeqOmegaX}
\end{align}
\end{small}
where the prime means derivative with respect to the redshift. 
To compare with the $\Lambda$CDM model, we use the expression 
\begin{align}
   E^{\Lambda CDM}(z)=\sqrt{\Omega_{r0}(1+z)^{4}+\Omega_{m0}(1+z)^{3}+\Omega_{\Lambda0}}, \label{HLambdaCDM}
\end{align}
where $\Omega_{r0}=2.469\times 10^{-5}h^{-2}(1+0.2271N_{\rm {eff}})$ and $\Omega_{\Lambda0}=1-\Omega_{r0}-\Omega_{m0}$. For these parameters we consider $H_{0}=67.4\;\rm km/s/Mpc$, $\Omega_{m0}=0.315$ and $N_{\rm {eff}}=2.99$ according to the Planck 2018 results \cite{Planck:2018vyg}.
For the constant values, we substitute the best-fit values found in \cite{Gonzalez:2023rsd} from SNe Ia+OHD data, say $\alpha= 0.416$, $\lambda= 2693\;\rm{Mpc}$, $H_{0}=67.4\;\rm km/s/Mpc$ and $l_0 = 10^{-34} \; \rm {cm}$, leading to  $\bar{\lambda}=181508/c, \bar{l}= 2.1843\times 10^{-57}/ c$. For the initial conditions, we choose the current time $z=0$ and $E(0)=1$ and $\Omega_m(0)=0.315$.
Therefore, we integrate the system 
\begin{footnotesize}
\begin{align}
    E^{\prime} & =\frac{E \Omega_m (4.69628\times 10^{-11} + 2.124 E^2 - 1.43497\times 10^{-113} E^4 \Omega_m)}{(1 + z) (E^2 + 3.13085\times 10^{-11} \Omega_m)}, \label{NumeqE}\\
    \Omega_m^{\prime} & =\frac{(E^2\Omega_m (3. + \Omega_m (-4.248 +  2.86993\times 10^{-113} E^2\Omega_m)))}{((1 +  z) (E^2 + 3.13085\times 10^{-11}\Omega_m))}. \label{NumeqOmegam}
\end{align}   
\end{footnotesize}
 \begin{figure}[t!]
    \centering
\includegraphics[scale=0.7]{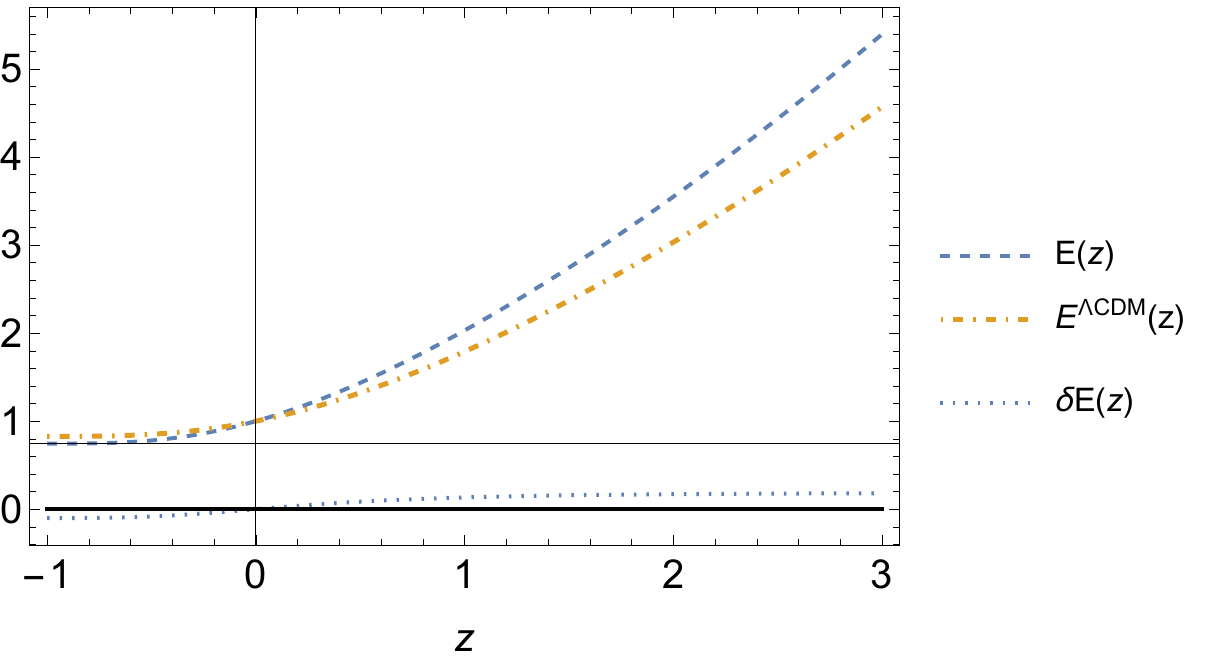}
    \caption{Numerical solution of $H/H_0(z)$ from \eqref{NumeqE}--\eqref{NumeqOmegam} compared with  $E^{\Lambda CDM}(z)$ from \eqref{HLambdaCDM} for the constant values $\bar{\lambda}=181508/c, \bar{l}= 2.1843\times 10^{-57}/ c$, and the initial conditions at the current time $z=0$ and $E(0)=1$ and $\Omega_m(0)=0.315$.}
    \label{fig:H[z]}
\end{figure}
We can compare the numerical solution of $H(z)$ from equations \eqref{NumeqE}--\eqref{NumeqOmegam} with $E^{\Lambda CDM}(z)$, which is calculated from the constant values $\bar{\lambda}=181508/c$ and $\bar{l}= 2.1843\times 10^{-57}/ c$. These values were obtained from the best-fit values found in \cite{Gonzalez:2023rsd} using SNe Ia+OHD data. The equation for $E^{\Lambda CDM}(z)$ is: $\sqrt{0.0000912556 (z+1)^4+0.315 (z+1)^3+0.684909}$. To begin our comparison, we use the initial conditions where $z=0$, $E(0)=1$, and $\ Omega_ (0)=0.315$. The results are displayed in Fig. \ref{fig:H[z]}, which shows the accuracy of our procedure in recovering $\Lambda CDM$ at late times, which is shown by comparing with zero the relative error 
\begin{equation}
    \delta E(z)= (E(z)- E^{\Lambda CDM}(z))/E^{\Lambda CDM}(z). 
\end{equation}

\section{Conclusions and Discussions}
\label{Sect:IV}
In the present work, we argue that cold dark matter and dark energy can be obtained from coupling the baryonic matter in terms of the gravitational field using a regular expression for the Yukawa gravitational potential. Such a quantum-corrected Yukawa-like gravitational potential leads to a long-range force, and it is characterized by the coupling parameter $\alpha$, the wavelength parameter $\lambda$ (which is related to the graviton mass), and a Planck length quantity $l_0$ that makes the potential regular for $r\rightarrow 0$. We have shown that due to the presence of gravitons, there is a contribution to the volume law entanglement entropy. That means that $\alpha$ results from the entanglement to the volume law entropy. 
For the galactic rotational curves and cosmological scales, the quantum effect is small hence $l_0 \rightarrow 0$, but $\alpha$ and $\lambda$, as we elaborated, can play an important role. 
For the velocity of stars in the outer part of galaxies, we obtain an expression for the velocity that tends to be constant in the outer part of the galaxy. We used Verlinde's entropic force interpretation, and we have obtained a new corrected Friedmann equation in which we can have two exceptional cases: the early Universe where the quantum effects, where quantum effects play an important role, and the late time universe, where apparent dark matter is essential. Very importantly, we obtain a new equation that relates the present density parameter of the dark matter $\Omega_{D}$ with the density parameter of the baryonic matter $\Omega_{B}$ ($z=0$). 

We presented a detailed dynamical analysis considering a single matter source and obtained the following results. When $-1<\omega<-\frac{1}{3}$, the system attractive lines such that $\rho = \rho_*$ and $H=\left(\frac{|\Gamma_2|}{ \left(1 + |\Gamma_1| \rho_*\right)}\right)^{\frac{1}{2}}$, where 
\begin{equation}
\rho_* = \frac{\left(\sqrt[3]{\Xi }-2 \sqrt[3]{\pi }   \sqrt[3]{G_{\rm eff}}\right)^2}{6 \sqrt[3]{\pi }
   \sqrt[3]{G_{\rm eff}} \sqrt[3]{\Xi } | \Gamma_{1}|},
   \end{equation} 
where $\Xi=8 \pi  G_{\rm eff} + 81 | \Gamma_1|  | \Gamma_2| -9
   \sqrt{| \Gamma_1|  | \Gamma_2| } \sqrt{81 | \Gamma_1|  | \Gamma_2| +16 \pi  G_{\rm eff}}$, representing a de Sitter solution. Similarly, when $ -\frac{1}{3}<\omega<\frac{1}{3}$, the lines corresponding to $\rho =\rho_*$ and $H=\frac{2 \sqrt{\frac{2 \pi }{3}} \sqrt{G_{\rm eff} \rho_*
   \left(y_c^2-1\right)}}{\sqrt{y_c^2-2}}$, where 
   \begin{equation}
   \rho_* :=  \small \frac{y_c^2}{2
   \Gamma_1} \pm \frac{\sqrt{G_{\rm eff} y_c^2 \left(4 \pi  G_{\rm eff}
   y_c^2 \left(y_c^2-1\right)^2+3 \Gamma_1
   \Gamma_2 \left(y_c^4-3
   y_c^2+2\right)\right)}}{4 \sqrt{\pi } \Gamma_1
   G_{\rm eff} \left(y_c^2-1\right)},
   \end{equation}
represents a late-time de Sitter solution. Finally, when $\omega>\frac{1}{3}$, the  system admits  lines of equilibrium points which corresponds to 
\begin{equation}
  \rho = \frac{1-2 y_c^2}{\Gamma_1}, H= \frac{\sqrt{|  \Gamma_{2}| }}{\sqrt{2}  y_c},  
\end{equation}, which are late-time de Sitter solutions. 
Assuming $ \Omega_B^{\Lambda CDM} \simeq 0.05$ along with $ \lambda \simeq  10^3 \,\, \rm Mpc$ and  $H_0 \simeq 70 \,\,\rm \rm km/s/Mpc$, along with $\alpha \simeq 0.04$, then from these estimations, we get the density parameter for the apparent dark matter and dark energy, as follows
\begin{eqnarray}\notag
\Omega_{\Lambda}^{\Lambda CDM}&=&(1+\alpha)\Omega_{\Lambda}=\frac{c^2 \alpha}{\lambda^2 H_0^2  (1+\alpha)  }\simeq 0.70,\\\notag
\Omega_D^{\Lambda CDM} &=&\sqrt{2 \Omega_{B}^{\Lambda CDM} \Omega_{\Lambda}^{\Lambda CDM} } \simeq 0.26,
\end{eqnarray}
which is quite remarkable. Further, if we take $\alpha \simeq 0.04$, on the other hand, we can obtain the graviton mass as follows
\begin{equation}\notag
   m_g=\frac{\hbar}{\lambda c} \simeq 1.14 \times 10^{-68}\, \rm kg \equiv 6.418\times 10^{-42} \rm {GeV}. 
\end{equation}
That clearly shows that dark matter results from coupling the baryonic matter via Yukawa potential; in particular, there is no dark matter particle in our model which means that dark matter is an apparent effect, i.e., a consequence of a modified gravity law. The cosmological constant can be estimated from the above values 
\begin{equation}\notag
\Lambda = \frac{3\,m_g^2 c^2\,\alpha}{\hbar^2\, (1+\alpha)^2 }\simeq 1.16 \times 10^{-52} \,\,\rm m^{-2}.
\end{equation}

In support of our view that dark matter is an emergent effect, we have found that there exists a correspondence with Verlinde's emergent gravity where dark matter is also viewed as an apparent effect; this correspondence is based on the following relation
\begin{equation}\notag
(c/\lambda H_0  (1+\alpha)) \times \sqrt{2\,\alpha\,\Omega_B } \rightarrow 2 \sqrt{\Omega_B}/\sqrt{3}.
\end{equation}
In other words,  on a fundamental level, gravity is an emergent phenomenon and effectively modifies the law of gravity precisely regarding Yukawa's potential. Finally, using the best-fit values along with the baryonic mass of the galaxy $ M\simeq 5 \times 10^{12} \times M_\odot\ $ and for the outer part of the galaxy, we can take $r \simeq 10^{20} $ m, we can see that 
\begin{equation}\notag
a_0= \frac{G M (r+\lambda)^2\alpha^2}{r^2\lambda^2}\simeq 10^{-10} \rm m/s^2.
\end{equation}
That shows that the theory reproduces the MOND phenomenology on galactic scales via the well-known acceleration of Milgrom.

\section*{CRediT authorship contribution statement}
 \textbf{Kimet Jusufi:} Conceptualization, Methodology, Formal analysis, Investigation, Writing – original draft, Writing – review \& editing, Supervision, Project administration. \textbf{Genly Leon:} Conceptualization, Writing – review \& editing, Visualization, Investigation, Formal analysis, Funding acquisition, Supervision, Project administration. \textbf{Alfredo D. Millano:}  Software, Validation, Formal analysis, Visualization, Investigation, Writing – review \& editing.

\section*{Declaration of competing interest}
The authors declare that they have no known competing financial interests or personal relationships that could have appeared to influence the work reported in this paper.

\section*{Data availability} No data is used in this article. 

\section*{Acknowledgements} G.L. was funded by Vicerrectoría de Investigación y Desarrollo Tecnológico (VRIDT) at Universidad Católica del Norte through Resolución VRIDT No. 040/2022, Resolución Vridt No. 054/2022, Resolución VRIDT No. 026/2023 and Resolución VRIDT No. 027/2023. He also thanks the support of Núcleo de Investigación Geometría Diferencial y Aplicaciones, Resolución Vridt No. 096/2022. ADM was supported by Agencia Nacional de Investigación y Desarrollo (ANID) Subdirección de Capital Humano/Doctorado Nacional/año 2020 folio 21200837, Gastos operacionales Proyecto de tesis/2022 folio 242220121.  We want to thank S. Arora and P.K. Sahoo for their discussions during the preparation of this work.

\bibliographystyle{apsrev4-1}

\bibliography{main.bib}

\end{document}